\documentclass{ifacconf}

\usepackage{graphicx}      
\usepackage{natbib}        	

\usepackage{amsmath} 	
\usepackage{amsfonts} 	
\usepackage{array} 			
\usepackage{enumerate}	

\usepackage{algorithm}
\usepackage{algpseudocode} 	

\usepackage{xcolor}			

\DeclareMathOperator*{\argmin}{arg\,min}
\newcommand*\diff{\mathop{}\!\mathrm{d}}

\makeatletter
\def\BState{\State\hskip-\ALG@thistlm}
\makeatother

\usepackage{eso-pic}

\begin{document}
\begin{frontmatter}

\AddToShipoutPictureBG*{%
	\AtPageUpperLeft{%
		\setlength\unitlength{1in}%
		\hspace*{\dimexpr0.5\paperwidth\relax}
		\makebox(0,-2)[c]{
			\parbox{\paperwidth}{ \centering
				Maurice Poot, On the Role of Models in Learning Control: Actor-Critic Iterative Learning Control, \\
				In {\em 21st IFAC World Congress}, Berlin, Germany, 2020 } }%
}}

\title{On the Role of Models in Learning Control: Actor-Critic Iterative Learning Control}


\author[First]{Maurice Poot} 
\author[Second]{Jim Portegies} 
\author[First]{Tom Oomen}

\address[First]{Control Systems Technology Group, Dept. of Mechanical Engineering, Eindhoven University of Technology, Eindhoven, The Netherlands. (m.m.poot@tue.nl).}
\address[Second]{CASA, Dept. of Mathematics and Computer Science, Eindhoven University of Technology, Eindhoven, The Netherlands.}

\begin{abstract}                
Learning from data of past tasks can substantially improve the accuracy of mechatronic systems. Often, for fast and safe learning a model of the system is required. The aim of this paper is to develop a model-free approach for fast and safe learning for mechatronic systems. The developed actor-critic iterative learning control (ACILC) framework uses a feedforward parameterization with basis functions. These basis functions encode implicit model knowledge and the actor-critic algorithm learns the feedforward parameters without explicitly using a model. Experimental results on a printer setup demonstrate that the developed ACILC framework is capable of achieving the same feedforward signal as preexisting model-based methods without using explicit model knowledge.

\end{abstract}


\end{frontmatter}

\section{Introduction}
Learning has large benefits for control applications, including high-tech mechatronic systems, by greatly improving the accuracy using data from past tasks. Commonly applied techniques, such as iterative learning control (ILC) (\cite{gunnarsson_norrlof_2001}) and repetitive control (\cite{RC}), learn from data to compensate for the error up to the limit of the reproducible behavior, see \cite{mikroniek}. However, these methods also exploit model knowledge for fast and safe learning. Obtaining a model of a system leads to user-intervention, which is not desired. In reinforcement learning (RL), many model-free learning techniques are developed that show promising convergence properties (\cite{Book:RL}; \cite{recht_2019}).

In ILC, the repetitive behavior of a system is exploited to learn a feedforward signal using a model and data of past tasks. Exceptional performance in terms of fast and safe learning requires the availability of a system model. Norm-optimal ILC (NOILC) is an important class in ILC where the optimal feedforward control signal is determined using a performance criterion, see \cite{gunnarsson_norrlof_2001} and \cite{amann_owens_rogers1996}. As a consequence of the assumption of repetitive tasks, the learned feedforward signal is only optimal for that specific task and the extrapolation to other tasks leads to performance deterioration (\cite{meulen_2008}). In \cite{phan_frueh_1996} basis functions for ILC are introduced to enhance the extrapolation properties by parameterizing the feedforward signal using basis functions in terms of the task. Additionally, the use of basis functions reduces the number of parameters that have to be learned in ILC and thus leads to a smaller computational burden, see \cite{zundert_2016}. Similarly, in \cite{rozario_oomen_2019} and \cite{banka_and_devasia} model-free approaches are presented to reduce the modeling requirement.

In contrast to the model-based approach of ILC, reinforcement learning (RL) is often used as a model-free and data-driven learning technique applied in robotics, games, and control (\cite{Book:RL}). Model-free methods are desired, as these limit the user-intervention and allow operation on different set of machines, or on machines of which it is difficult to find a model. In RL, optimal control problems described by a Markov decision process (MDP) (\cite{Book:RL}) are solved without explicitly using a model. For mechatronic systems described by continuous MDPs, discretization could be applied to allow the use of tabular methods, however, these methods would suffer from the curse of dimensionality to allow accurate control. To achieve model-free learning from experience in continuous MDPs, approximate RL techniques can be employed (\cite{recht_2019}), such as the actor-critic algorithm described in \cite{busoniu_2018} and \cite{grondman_et_al_2012}. Related attempts to investigate model aspects in control and RL include \cite{turchetta}, \cite{berkenkamp_et_al}, and \cite{berkenkamp_and_schoellig}.


Although ILC methods often have exceptional performance and fast and safe convergence properties, these methods require the availability of a model of the system. The aim of this paper is to investigate model-based and model-free learning for mechatronic systems and to develop a model-free approach to learn the optimal feedforward parameters in the ILC framework using RL. The main contribution is the developed actor-critic iterative learning control (ACILC) framework. 




This paper is structured as follows. In Section \ref{sec:problem}, the learning scheme for a system with basis functions is given and the problem is formulated. In Section \ref{sec:ACILC} the ACILC framework is formulated using the actor-critic algorithm. In Section \ref{sec:ILC}, the model-based norm-optimal ILC solution is explained. In Section \ref{sec:ex}, experimental results on a printer setup are presented. In Section \ref{sec:conclusions}, conclusions are given.


\section{Problem formulation}
\label{sec:problem}
In this section, the learning control problem with basis functions is formulated.

\begin{figure}[t!]
\centering
\includegraphics[width=2.5in]{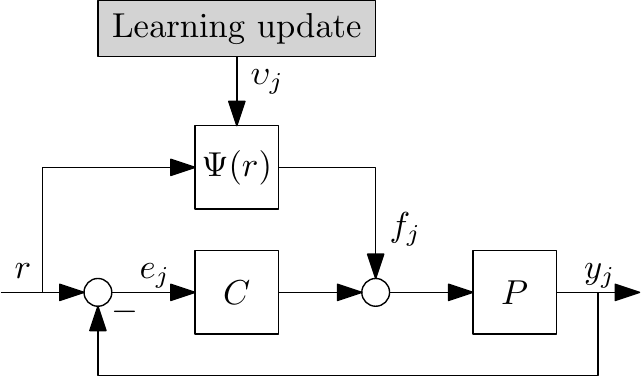}
\caption{Control scheme with basis functions and learning update structure.}
\label{fig:control_scheme_ILCBF_learning_update}
\end{figure}

Consider the closed-loop system shown in Figure \ref{fig:control_scheme_ILCBF_learning_update}. The system and feedback controller are denoted by $P$ and $C$, respectively. Both are considered to be linear time-invariant (LTI), causal, and single-input single-output. An experiment, also called trial, has index $j$, length $N \in \mathbb{Z}^+$, reference $r \in \mathbb{R}^N$ and measured output $y_j \in \mathbb{R}^N$. From \ref{fig:control_scheme_ILCBF_learning_update} it follows that the error at trial $j$ in this lifted representation is given by
\begin{equation}
e_j = Sr-SPf_j,
\label{eq:ILC_e_j}
\end{equation}
where $S = (I+PC)^{-1}$ is the sensitivity and $J=SP$ is the impulse response matrix of the process sensitivity $SP$. The controller $C$ is assumed to be internally stabilizing, i.e., $S, SP, SC, SPC \in \mathcal{H}_\infty$. The feedforward signal is parameterized by a user-defined basis function $\Psi$, i.e.,
\begin{equation}
f_j = \Psi \upsilon_j,
\label{eq:ILC_BF_f_j}
\end{equation}
with $\Psi~=~[ \psi_1,  \psi_2, \dots, \psi_m ]~\in ~\mathbb{R}^{N\times m}$ and parameters $\upsilon \in \mathbb{R}^m$, and $m$ is the number of basis functions. By choosing $\Psi$ as a function of the reference $r$, the error and the learned feedforward signal become independent of the reference, see Section \ref{sec:ILC} and \cite{wijdeven_bosgra_2010}. Note that if $\Psi=I$, then standard ILC is recovered for which the extrapolation to other reference trajectory leads to performance deterioration (\cite{meulen_2008}). Additionally, if the basis is chosen as $\Psi(r)\upsilon_j = P^{-1}$, optimal tracking, i.e., $e_j=0$, can be achieved, see \cite{bolder_oomen_2015} for proof. 

The error propagation from trial $j$ to trial $j+1$ follows by evaluating \eqref{eq:ILC_e_j} for $j$ and $j+1$ and subsequently eliminating $Sr$, leading to
\begin{align}
e_{j+1} &= Sr - J \Psi(r) \upsilon_{j+1} \nonumber \\
&= e_j - J \Psi(r) ( \upsilon_{j+1} - \upsilon_j ).
\label{eq:ILC_BF_e_j+1}
\end{align}
The objective is to minimize the error $e_{j+1}$ by selecting the feedforward parameters $\upsilon_{j+1}$ based on the measured signals $e_j$ and $\upsilon_j$.

In case a model of $J$ is available, then $\upsilon_{j+1}$ can be computed by minimizing $e_{j+1}$ using a performance criterion, which is called norm-optimal ILC, see Section \ref{sec:ILC}. Although fast convergence can be achieved using ILC, these methods require an explicit model of $J$ for fast and safe learning. The requirements on a model increase the amount of user-interventions necessary. Many applications have machine-to-machine differences, or have models which are hard to obtain.

The problem that is considered in this paper is to
\begin{enumerate} 
\item learn $\upsilon_{j+1}$ from the measured signals $\upsilon_j$ and $e_j$ without explicitly using a model;
\item allow incorporation of implicit model knowledge;
\item and enable extrapolation to different tasks.
\end{enumerate}

In the next section, a new approach is presented that learns the feedforward parameters from data without explicitly using a model, which is the main contribution of this paper.

\section{ACILC}
\label{sec:ACILC}
In this section, the ACILC approach that addresses the problem of Section \ref{sec:problem} is presented. In Section \ref{sec:MDP}, the Markov decision process (MDP) for the described system is formulated which is used to describe the optimal control problem. In Section \ref{sec:AC} the actor-critic learning technique is presented to solve the optimal control problem without using an explicit model of the system.


\subsection{Learning setup}
\label{sec:MDP}
The general idea of reinforcement learning (RL) is solving optimal control problems that are formulated as MDPs (\cite{Book:RL}), which is a way to describe an optimal control problem for RL.


The closed-loop system of \eqref{eq:ILC_BF_e_j+1} with basis functions is considered and described as a Markov decision problem. The MDP is defined by the tuple ($\mathcal{X}, \mathcal{U}, \mathcal{T},  c, \gamma$), where
\begin{itemize}

\item $\mathcal{X}$ is the set of states. The state at trial $j$ is defined by the error $e_j \in \mathbb{R}^N$ and the feedforward parameters $\upsilon_j \in \mathbb{R}^m$, i.e., the available data at trial $j$.

\item $\mathcal{U}$ is the set of actions. The action for trial $j$ is defined as the feedforward parameters $\upsilon_{j+1} \in \mathbb{R}^m$, i.e., the action that is selected in state $\{e_j,\upsilon_j\}$.

\item $\mathcal{T}(e_j,\upsilon_j,\upsilon_{j+1})$, is the transition function given in \eqref{eq:ILC_BF_e_j+1}, where $\mathcal{T}: \mathcal{X} \times \mathcal{U} \rightarrow \mathcal{X}$. It describes the transition from state $\{e_j,\upsilon_j\}$ to the next state $e_{j+1}$ after applying action $\upsilon_{j+1}$

\item $c(e_j,\upsilon_j,\upsilon_{j+1})$ is the cost function in $\mathbb{R}$, where $c: \mathcal{X} \times \mathcal{U} \rightarrow \mathbb{R}$. It describes the immediate cost of the state $\{e_j,\upsilon_j\}$ and applying action $\upsilon_{j+1}$.

\item $\gamma$ is the discount factor in (0,1]. It represents the difference in importance between immediate cost and future cost.

\end{itemize}

The optimization problem described by the MDP is as follows. The agent chooses actions according to a policy $\pi$, where $\pi: \mathcal{X} \rightarrow \mathcal{U}$, mapping the state to action. The goal of the agent is to find the policy $\pi$ which minimizes the infinite horizon discounted cost, which is the sum of the accumulated cost described by
\begin{equation}
\sum^\infty_{j=0} \gamma^j c(e_j,\upsilon_j,\upsilon_{j+1}),
\label{eq:RL_ILC_return}
\end{equation}
where the cost of trial $j$ is described by the quadratic function, 
\begin{equation}
c(e_j,\upsilon_j,\upsilon_{j+1}) = \parallel e_j \parallel^2_{W_e} + \parallel \upsilon_{j+1} \parallel^2_{W_{\upsilon}} + \parallel \upsilon_{j+1} - \upsilon_j \parallel^2_{W_{\Delta \upsilon}}
\label{eq:ACILC_cost}
\end{equation}
where $\parallel x \parallel_W = x^\top W x$. Note that this cost is a weighting of the state in trial $j$ and the action selected in that state, $\upsilon_{j+1}$. 
The objective is to learn a policy to determine the action $\upsilon_{j+1}$ from the experiment data $e_j$ and $\upsilon_j$, which is described by the state feedback
\begin{equation}
\upsilon_{j+1} = \pi(e_j, \upsilon_j).
\label{eq:RL_ILC_general_policy}
\end{equation}
The cost, or value, of a policy $\pi$ from an initial state $\{e_0,\upsilon_0\}$ is the discounted return
\begin{equation}
V^\pi(e_0,\upsilon_0) = \sum^\infty_{j=0} \gamma^j c\big( e_j,\upsilon_j,\pi(e_j, \upsilon_j) \big),
\end{equation}
where $\upsilon_{j+1}$ is given in \eqref{eq:RL_ILC_general_policy}. The optimal value of a state $\{e_j,\upsilon_j\}$ is described by the Bellman optimality equation, as described by
\begin{equation}
V^*(e_j,\upsilon_j) = \min_{\upsilon_{j+1}} \Big[ c(e_j,\upsilon_j,\upsilon_{j+1}) + \gamma V^*\big( \mathcal{T}(e_j,\upsilon_j,\upsilon_{j+1}) \big)  \Big].
\label{eq:RL_ILC_Vstar}
\end{equation}

The solution of the Bellman optimality equation can be computed using optimal control when the model is available, as shown in Section \ref{sec:ILC}. However, in the absence of a model a reinforcement learning technique can be employed. Since the state and action spaces of the MDP are continuous, approximate RL algorithms, such as the actor-critic algorithm, are necessary \cite[p. 273]{Book:RL}, \cite[p. 9]{busoniu_2018}, as shown next.


\subsection{Actor-critic learning}
\label{sec:AC}
The actor-critic algorithm for the solution of the optimal control problem is presented here. Herein, a \textit{critic} estimates the value of a state using a value function approximation and a \textit{actor} estimates the optimal approximated policy.



\subsubsection{Critic}
The objective of the critic is to estimate the value of a state using a value function approximation. The value of a state is described by a set of feature vectors along which generalizations of the state space are appropriate and the weights of these features are adjusted to obtain a better estimate of the value of the state. The value function approximation is chosen to be linear in weights and as a function of the projection of the state instead of the state $\{e_j,\upsilon_j\}$ itself. Thus, the value $V^\pi(e_j,\upsilon_j) : \mathcal{X} \times \mathbb{R}^{m_w} \rightarrow \mathbb{R}$, for a policy $\pi$ is approximated by
\begin{equation}
V^\pi(e_j,\upsilon_j) \approx \hat{V}^\pi(x_j,w_j) = w_j^\top \phi_w(x_j)
\label{eq:RL_ILC_V_approx}
\end{equation}
where $w \in \mathbb{R}^{m_w}$ denotes the parameter vector, $\phi_w(x_j)$ the feature vector, and $x_j \in \mathbb{R}^m$ the correlation of the error $e_j \in \mathbb{R}^N$ with the basis $\Psi(r)\in \mathbb{R}^{N \times m}$, given by
\begin{equation}
x_j = \Psi^\top(r) e_j.
\label{eq:RL_ILC_x_projection}
\end{equation}
The feature vector is simply chosen equal to the projection of the error
\begin{equation}
\quad \phi_w (x_j) = x_j.
\end{equation}
By using the projection of the error $x_j$ as the state of the value function instead of $\{e_j,\upsilon_j\}$, generalizations are made along the reduced state space, which significantly reduces the size of the feature vector and the parameter vector. Additionally, the choice for the basis functions allows the incorporation of prior knowledge and enables extrapolation to different motion tasks, making the value function approximate invariant under changes of the reference signal.


The critic parameters are adjusted in the direction of reducing the estimation error, which is described by the difference between the right-hand side and the left-hand sides of the Bellman equation \eqref{eq:RL_ILC_Vstar}, i.e.,
\begin{equation}
\delta_j = c(e_j,\upsilon_j,\upsilon_{j+1}) + \gamma \hat{V}(x_{j+1},w_j) - \hat{V}(x_j,w_j),
\label{eq:RL_ILC_TD_error}
\end{equation}
where $\delta_j$ is called the temporal-difference error. The direction in which to adjust the weights to reduce the estimation error is determined by the gradient of the value function approximate, as explained in \cite[p.~164]{Book:RL}. The update of the critic parameters is given by
\begin{align}
w_{j+1} &= w_j + \alpha_w \delta_j \nabla_w \hat{V}(x_j,w_j) \nonumber \\
&= w_j + \alpha_w \delta_j \phi_w(x_j),
\label{eq:ACILC_critic_update}
\end{align}
where $\alpha_w$ is the learning rate, $\delta_j$ is the temporal-difference error as a scalar metric for how accurate the current estimate is, and $\phi_w(x_j)$ the feature vector that describes the direction in which to change the parameters. Note that the approximation is linear, hence the gradient of the value function is the feature vector.

The critic update can be performed online, at every trial $j$, and every encountered state contributes to minimizing the estimation error. The update of the parameters and the usage of the temporal-difference error follows the work of \cite[p. 161]{Book:RL}. However, the usage of a basis function to reduce the size of the state is new. Next, the actor is described.


\subsubsection{Actor}
The goal of the actor is to find the optimal policy that solves \eqref{eq:RL_ILC_Vstar}. By using a parameterized function as the policy, continuous actions $\upsilon_{j+1}$ can be selected. The policy is optimized by adjusting the parameters using the value function estimate of the critic.



To facilitate the selection of real-valued actions $\upsilon_{j+1}$, the policy is approximated by a probability density function over the continuous action space $\upsilon_{j+1} \in \mathcal{U}$ for each $x_j \in \mathcal{X}$ and $\theta \in \mathbb{R}^{m_\theta \times m_\theta}$. The approximate policy $\hat{\pi}(\upsilon|x,\theta)$ is given by
\begin{equation}
\hat{\pi}(\upsilon_{j+1} |x_j,\theta_j) = \frac{1}{\sigma\sqrt{2\pi}} \exp \left( -\frac{ \big(\upsilon_{j+1} -\mu(x_j,\theta_j)\big)^2}{2\sigma^2} \right),
\label{eq:RL_gaussian_policy}
\end{equation} 
where $\mu(x_j,\theta_j): \mathcal{X} \times \mathbb{R}^{m_\theta} \rightarrow \mathbb{R}^{m_\theta}$ is the mean which is parameterized by $\theta$ and $\sigma^2 \in \mathbb{R}^+$ is the variance which is constant with respect to the parameters. Note that the state projection $x_j$ is chosen as the state for the same reasoning as mentioned in the previous section. The mean of this policy approximation is defined by a linear function
\begin{equation}
\mu(x_j,\theta_j) = \theta_j^\top \phi_\theta(x_j),
\label{eq:ACILC_policy}
\end{equation}
where $\theta_j$ are the actor parameters and $\phi_\theta(x_j)$ are feature vectors as a function of the projected error of \eqref{eq:RL_ILC_x_projection}. Again, the feature vectors are chosen as the projected state, i.e.,
\begin{equation}
\phi_\theta (x_j) = x_j.
\end{equation}

Using this policy approximation, the real-valued actions $\upsilon_{j+1}$ can be drawn from the Gaussian distribution using
\begin{equation}
\upsilon_{j+1} \sim \mathcal{N}\big( \mu(x_j,\theta_j),\sigma^2 I \big)
\label{eq:ACILC_draw_action}
\end{equation}
where $\upsilon_{j+1}$ is drawn with mean $\mu~\in~\mathbb{R}^{m_\theta}$ and variance $\sigma^2~\in~\mathbb{R}^{+}$. By deviating from the mean, exploration is achieved which is necessary for learning. The variance should decay with respect to $j$ to allow for exploitation, but should not go to zero to ensure continuous learning in the case of changing system parameters.



Next, the policy parameters are adjusted such that the infinite sum of the accumulated cost, i.e., true value function, is minimized. The gradient of the true value function with respect to the policy parameters describes the direction of steepest descent for the parameters, and could be used to update the parameters. The gradient can not be computed directly since the true value function is unknown and depends on the visited states and the selected actions, which both depend on the policy parameters. However, using the policy gradient theorem (\cite{sutton_2000}), the gradient is sampled, i.e., every visited state and selected action contribute to the estimation of the gradient. The update of the parameters is given by
\begin{equation}
\theta_{j+1} = \theta_j - \alpha_\theta \delta_j \nabla_\theta \log \hat{\pi}(\upsilon_{j+1}|x_j,\theta_j)
\label{eq:ACILC_actor_update}
\end{equation}
where $\alpha_\theta$ is the learning rate, $\delta_j$ is the temporal-difference error of the critic \eqref{eq:RL_ILC_TD_error}, and $\nabla_\theta \log \hat{\pi}(\upsilon_{j+1}|x_j,\theta_j)$ is the so-called likelihood ratio. The latter is an analytic expression for sampling the gradient and is computed from \eqref{eq:RL_gaussian_policy}, leading to
\begin{equation}
\nabla_\theta \log \hat{\pi}(\upsilon_{j+1}|x_j,\theta_j) = \frac{\upsilon_{j+1}-\mu(x_j,\theta_j)}{\sigma^2}\phi_\theta(x_j).
\end{equation}

The actor allows for selection of continuous actions at every trial $j$ using a parameterized policy, see \cite[p. 277]{Book:RL}. The optimization of the policy parameters uses the policy gradient theorem (\cite{sutton_2000}) and the actor-critic framework, see \cite{grondman_et_al_2012} and \cite[p. 265]{Book:RL}. Next, the complete actor-critic algorithm is described.

\subsubsection{Algorithm}
The complete actor-critic algorithm is described in Algorithm \ref{alg:ACILC} and describes the learning update in Figure \ref{fig:control_scheme_ILCBF_learning_update}.
\begin{algorithm}
\caption{ACILC}
\label{alg:ACILC}
\begin{algorithmic}[1]
\BState Set $\alpha_w$, $\alpha_\theta$, $\sigma^2$, $\phi_w$, $\phi_\theta$, and 
\BState Choose $\Psi(r)$, $W_e$, $W_\upsilon$, and $W_{\Delta \upsilon}$
\BState Initialize $w_0$, $\theta_0$, and $x_0$
\For {$j = 0,1,2,3,...$}
\State Draw action $\upsilon_{j}$
\State Perform experiment, transition to next state $e_{j+1}$
\State Calculate $x_{j+1}$ and $\delta_{j+1}$
\State Update critic parameters $\theta_{j+1}$
\State Update actor parameters $w_{j+1}$
\EndFor
\end{algorithmic}
\end{algorithm}
For convergence of the actor-critic algorithm, the learning rates $\alpha_w$ and $\alpha_\theta$ should be chosen such that the learning speed is reduced with respect to the trial number, see \cite{grondman_et_al_2012} for conditions. To allow for continuous learning, for example in the case of changing system parameters due to wear and tear, the value of learning rates should not go to zero.

To conclude, ACILC address the problem of Section \ref{sec:problem} by learning the optimal feedforward parameters from experience. Using basis functions in the value function approximation, prior model knowledge can be incorporated and extrapolation capabilities are enabled. In the next section, the existing model-based approach using norm-optimal ILC is explained, after which experimental results are shown to compare the model-free and model-based methods.

\section{Norm-optimal ILC with basis functions}
\label{sec:ILC}


In this section, norm-optimal iterative learning control (NOILC) is presented for comparison to ACILC. NOILC is a model-based approach for deriving the feedforward parameters $\upsilon_{j+1}$ by optimizing a performance criterion (\cite{gunnarsson_norrlof_2001}).

The general performance criterion for norm-optimal ILC with basis functions is given by
\begin{equation}
\mathcal{J}(\upsilon_{j+1}) = \parallel e_{j+1} \parallel^2_{W_e} + \parallel \upsilon_{j+1} \parallel^2_{W_\upsilon} + \parallel \upsilon_{j+1}-\upsilon_{j} \parallel^2_{W_{\Delta \upsilon}}
\label{eq:NOILC_BF_perf_crit}
\end{equation}
with user-defined weighting $W_e \succ 0$, $W_\upsilon, W_{\Delta \upsilon} \succeq 0$, and $e_{j+1}$ given by \eqref{eq:ILC_BF_e_j+1}. Here, the optimization description considers the error $e_{j+1}$ of the next trial, which is described using the model in \eqref{eq:ILC_BF_e_j+1}. The optimal feedforward parameters $\upsilon^*_{j+1}$, that minimizes the performance criterion \eqref{eq:NOILC_BF_perf_crit}, are defined by
\begin{equation}
\upsilon^*_{j+1} = \argmin_{\upsilon_{j+1}} \mathcal{J}(\upsilon_{j+1}),
\label{eq:NOILC_BF_argmin}
\end{equation}
and can be computed analytically since \eqref{eq:NOILC_BF_perf_crit} is quadratic and \eqref{eq:ILC_BF_e_j+1} is linear in $\upsilon_{j+1}$. Note that this optimization description only considers the cost of the next trial. Hence, \ref{eq:NOILC_BF_argmin} is the one-step-look-ahead solution of \eqref{eq:RL_ILC_Vstar}.

For a $\Psi(r)$ with full column rank, the solution directly follows from the necessary condition of optimality $\frac{\partial \mathcal{J}}{\partial f_{j+1}}~=~0$, and is given by
\begin{equation}
\upsilon^*_{j+1} = Q \upsilon_j + L e_j
\label{eq:NOILC_BF_ups_star}
\end{equation}
with
\begin{align}
Q = &\left[ \Psi^\top \left(  J^\top W_e J  +  W_f  +  W_{\Delta f}  \right) \Psi \right]^{-1} \nonumber \\
 &\cdot \Psi^\top \left(  J^\top W_e J + W_{\Delta f} \right) \Psi, \nonumber \\
L = &\left[ \Psi^\top \left(  J^\top W_e J  +  W_f  +  W_{\Delta f}  \right) \Psi \right]^{-1} \Psi^\top J^\top W_e,
\label{eq:NOILC_BF_QandL}
\end{align}
where $Q \in \mathbb{R}^{m \times m}$ and $L \in \mathbb{R}^{m \times N}$ are, respectively, the robustness matrix and learning matrix.
%
%
The solution of \eqref{eq:NOILC_BF_ups_star} describes the model-based learning update of Figure \ref{fig:control_scheme_ILCBF_learning_update}, and is monotonic convergent with respect to the 2-norm of the error if
\begin{equation}
\bar{\sigma}(Q - LJ\Psi) < 1,
\end{equation}
and is guaranteed if $\Psi^\top \left(  J^\top W_e J  +  W_\upsilon +  W_{\Delta \upsilon}  \right) \Psi \succ 0$. In the case that $J$ is singular, monotonic convergence can still be guaranteed by setting the weight $W_\upsilon \succ 0$ or by exploiting the freedom in $\Psi$. By choosing $W_{\Delta \upsilon} \succeq 0$ attenuation of trial-varying disturbances is achieved at the cost of convergence speed (\cite{oomen_rojas_2017}).

In conclusion, a model-based approach is presented to derive the optimal feedforward parameters. In the next section, the new model-free method is compared to the model-based method using experiments.
\section{Experiments}
\label{sec:ex}
In this section, the effectiveness of ACILC is experimentally demonstrated, revealing the model-free performance in comparison to preexisting NOILC with basis functions.

\subsection{Experimental setup}

\begin{figure}[!t]
\centering
\setlength{\fboxsep}{0pt}
\fbox{\includegraphics[width=0.49\textwidth]{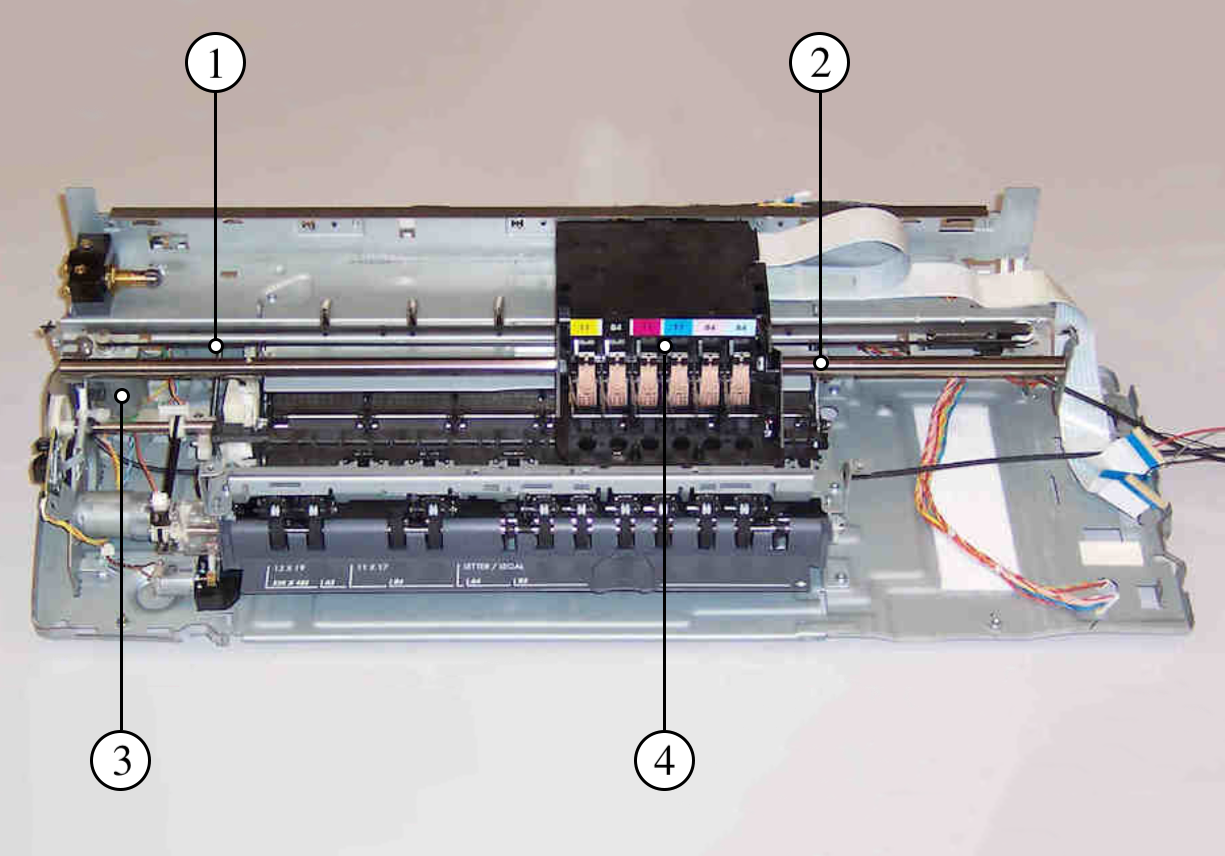}}
\caption{Printer motion system with: (1) drive belt, (2) slide guide, (3) motor, and (4) carriage with printhead.}
\label{fig:printer}
\end{figure}

The experimental printer setup is shown in Figure \ref{fig:printer}. The system consists of a current controlled motor with input range of $\pm2.5$V driving a carriage that slides along a guide. The carriage position is measured by a 600 counts per inch optical encoder. The drive belt that connects the motor to the carriage has finite stiffness.
A discrete-time parametric model is obtained from frequency response measurements. The obtained model does not correspond to this specific unit per se, since different setups exist. The plant and feedback controller are discretized using a first-order-hold with sample time $1$~ms, and are give by
\begin{align}
P(z) &= \frac{  1 \times 10^{-8}  \big( 24.24 z^3 + 130.3 z^2 + 32.95 z - 8.486 \big) }{z^5 - 3.761 z^4 + 5.438 z^3 - 3.593 z^2 + 0.9157 z} \nonumber \\
C(z) &= \frac{108.6 z^3 + 112.9 z^2 - 100 z - 104.3}{z^3 - 0.6499 z^2 - 0.9465 z + 0.7035}.
\end{align}


%

The reference trajectory $r$ consists of two third-order point-to-point motions with a short rest in between.
The experiments are repeated for 40 trials, starting at trial $j=0$ with no feedforward, i.e., $\upsilon_0 = 0$.


\subsection{Design aspects}
The error of the system is expected to be dominated by friction and acceleration errors. Using this prior knowledge, the basis for both methods are chosen as
\begin{equation}
\Psi(r) = \begin{bmatrix}
\frac{\diff^2}{\diff t^2} r & \frac{\diff}{\diff t} r
\end{bmatrix}.
\label{eq:example_basis}
\end{equation}
Additionally, the weighting matrices for the performance criterions \eqref{eq:ACILC_cost} and \eqref{eq:NOILC_BF_perf_crit} are designed for performance and not robustness, i.e.,
\begin{equation}
W_e = I \cdot 10^{6}, \quad W_\upsilon = I \cdot 10^{-6}, \quad \text{and} \quad W_{\Delta \upsilon} = 0.
\label{eq:example_weighting}
\end{equation}
Since $J$ is not full rank due to delays in the system, $W_\upsilon \succ0$ is required for monotonic convergence in the 2-norm of the error for NOILC.
The weighting for ACILC is chosen identical to allow direct comparison of the cost.

The NOILC with basis functions is synthesized using the system model as described in Section \ref{sec:ILC}. The ACILC algorithm is updated online after every trial $j$, as described in Algorithm \ref{alg:ACILC}. The learning rates $\alpha_\theta$ and $\alpha_w$, and the exploration parameter $\sigma$ are decaying exponentially with respect to the trial number $j$.




\subsection{Results}

\begin{figure}[!t]
\centering
\includegraphics[width=0.49\textwidth]{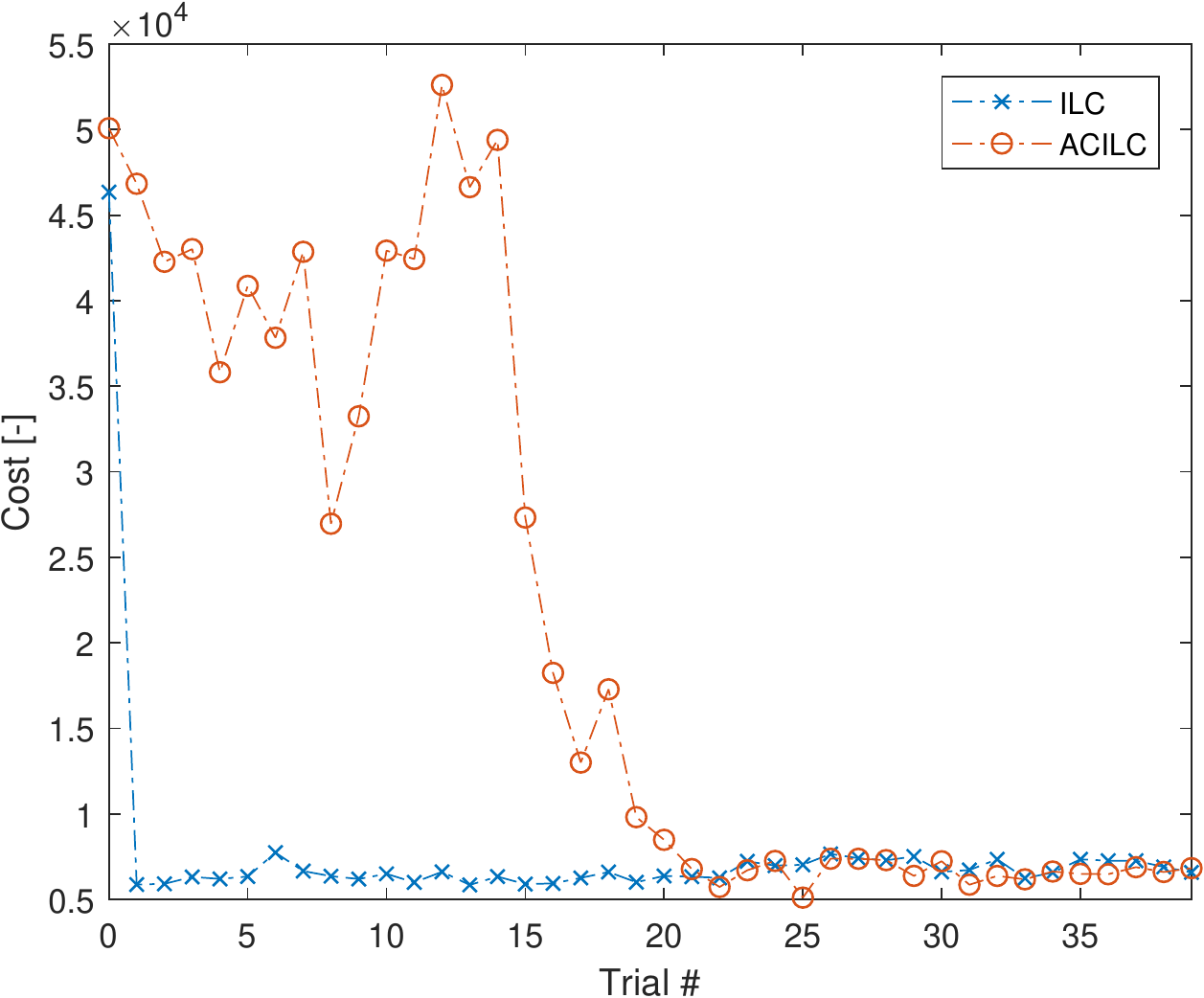}
\caption{Cost per trial for the ILC with basis functions and ACILC experiments. Note that the final cost of both methods is comparable.}
\label{fig:YES_cost}
\end{figure}

\begin{figure}[!t]
\centering
\includegraphics[width=0.49\textwidth]{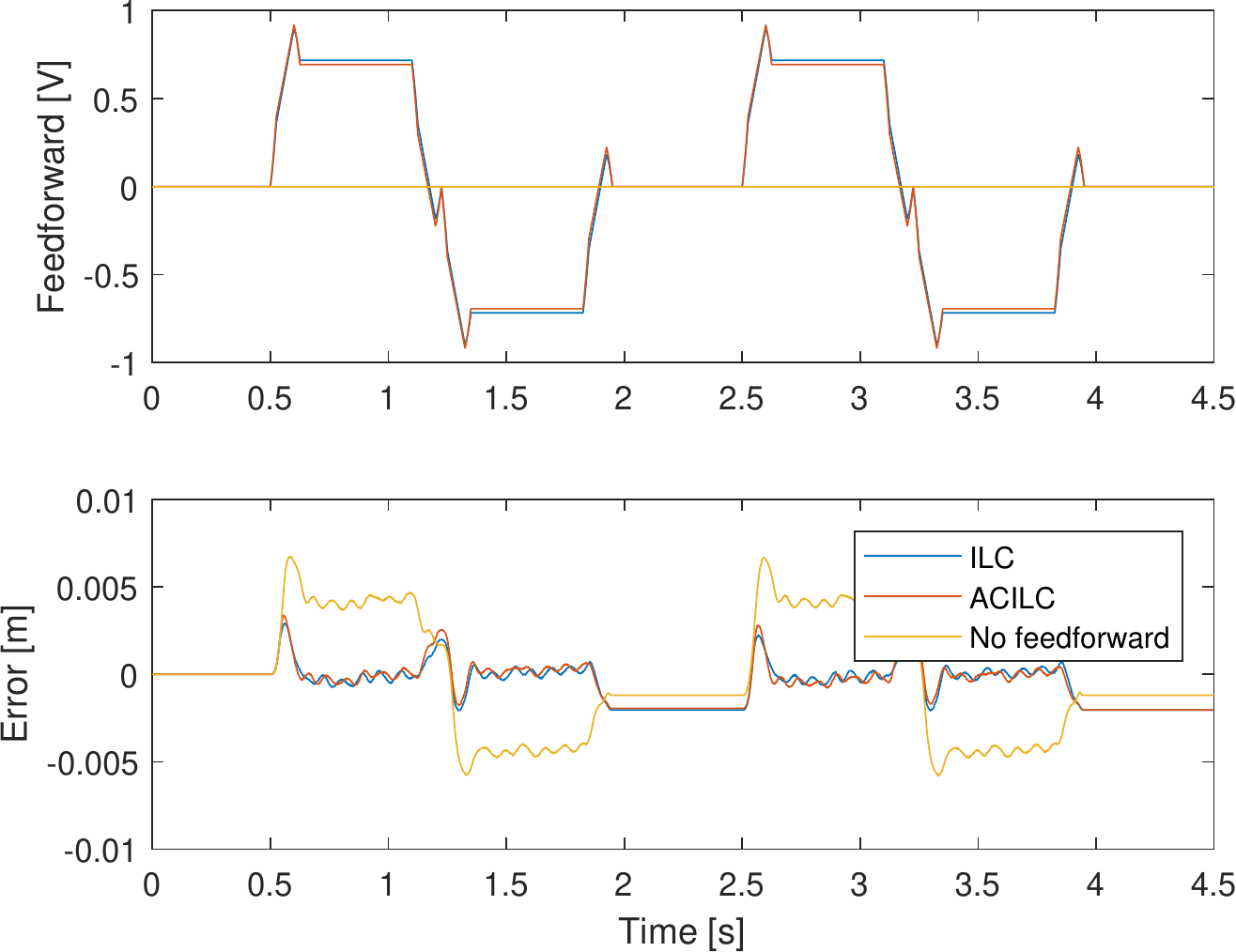}
\caption{Feedforward signal and measured error signal at trial $40$ for the ILC with basis functions (blue) and ACILC (red). Both methods compensate for the errors due to friction and acceleration.}
\label{fig:YES_fande_plus_initial}
\end{figure}

The experiments of ACILC and ILC with basis functions are performed on the same system and the key result is the cost per trial, shown in Figure \ref{fig:YES_cost}. The actor-critic cost varies significantly in the first few trials due to the variance in the Gaussian policy necessary for exploration. After trial 15 the parameters of the critic are converged and a good representation of the value function is learned, thereafter, the actor parameters converge, resulting in a lower cost. The ACILC algorithm eventually achieves the same final cost as the model-based ILC method.

The resulting feedforward signal for the last trial, and the corresponding error signal are depicted in Figure \ref{fig:YES_fande_plus_initial}.
It shows that the feedforward signal is able to compensate for the mass acceleration and viscous friction errors and that ACILC produces the same feedforward and error signal as the ILC with basis functions without the use of an explicit model.

\section{Conclusion and outlook}
\label{sec:conclusions}

Actor-critic iterative learning control (ACILC) is a new data efficient approach to learn a feedforward signal from data using limited prior model knowledge. The model-free framework that is developed uses the actor-critic reinforcement learning algorithm to learn the optimal feedforward parameters without requiring an explicit model. The experimental results demonstrate that the ACILC framework is capable of achieving the same feedforward signal as NOILC with basis functions with very limited and implicit model knowledge, while at the same time still achieving good convergence rates.



Future research focuses on scaling and tuning of the actor-critic parameters to increase performance and robustness. Secondly, the selection of basis functions from implicit model knowledge or data to increase performance and on obtaining similar performance without the use of any prior knowledge is investigated. Thirdly, the capabilities of ACILC to cope with different cases, such as extrapolation to different motion tasks and changing system parameters are planned for experimental validation. Finally, the applicability for different systems, including non-LTI systems, is investigated to extend the usability of the algorithm.



\begin{ack}
This work is supported by ASM Pacific Technology and NWO-VIDI nr 15698.
\end{ack}


\begin{thebibliography}{xx} 

\bibitem[Amann et al., 1996]{amann_owens_rogers1996} Amann, N., Owens, D., and Rogers, E. (1996). Iterative learning control using optimal feedback and feedforward actions. \textit{Int. Journal of Control}, 65(2), 277-293.

\bibitem[Banka and Devasia, 2018]{banka_and_devasia} Banka, N. and Devasia, S. (2018). Application of iterative machine learning for output tracking with magnetic soft actuators. \textit{IEEE/ASME Transactions on Mechatronics}, 23, 2186-2195.

\bibitem[Berkenkamp et al., 2017]{berkenkamp_et_al} Berkenkamp, F., Turchetta, M., Schoellig, A., and Krause, A. (2017). Safe model-based reinforcement learning with stability guarantees. \textit{Proc. of the 31st Int. Conf. on Neural Information Processing Systems (NIPS)}, 908-919.


\bibitem[Berkenkamp and Schoellig, 2015]{berkenkamp_and_schoellig} Berkenkamp, F. and Schoellig, A. (2015). Safe and robust learning control with gaussian processes. \textit{2015 European Control Conference (ECC)}, 2496-2501.


\bibitem[Bolder and Oomen, 2015]{bolder_oomen_2015} Bolder, J. and Oomen, T. (2015). Rational basis functions in iterative learning control - with experimental verification on a motion system. \textit{IEEE Trans. Control Syst. Technol.}, 23(2), 722-729.

\bibitem[Bu\c{s}oniu et al., 2018]{busoniu_2018} Bu\c{s}oniu, L., de Bruin, T., Toli\'{c}, D., Kober, J., and Palunko, I. (2018). Reinforcement learning for control: performance, stability, and deep approximators. \textit{Annual Reviews in Control}, 46, 8-28.

\bibitem[Grondman et al., 2012]{grondman_et_al_2012} Grondman, I., Bu\c{s}oniu, L., Lopes, G., and Babu\u{s}ka, R. (2012). A survey of actor-critic reinforcement learning: standard and natural policy gradients. \textit{IEEE Trans. on Systems, Man, and Cybernetics, Part C (Appl. and Rev.)}, 42(6), 1291-1307.

\bibitem[Gunnarsson and Norrl\"{o}f, 2001]{gunnarsson_norrlof_2001} Gunnarsson, S. and Norrl\"{o}f, M. (2001). On the design of ILC algorithms using optimization. \textit{Automatica}, 37(12), 2011-2016.

\bibitem[Hara et. al., 1988]{RC} Hara, S., Yamamoto, Y., Omata, T., and Nakano, M. (1988). Repetitive control system: a new type servo system for periodic exogenous signals. In \textit{IEEE Trans. on Automatic Control}, 33(7), 659-668.

\bibitem[Meulen et al., 2008]{meulen_2008} van der Meulen, S., Tousain, R., and Bosgra, O. (2008). Fixed structure feedforward controller design exploiting iterative trials: application to a wafer stage and a desktop printer. \textit{Journal of Dynamic Systems, Measurement and Control: Transactions of the ASME}, 130(5), 1-16.

\bibitem[Nemec et al., 2017]{nemec_2017} Nemec, B., Simoni\v{c}, M., Likar, N., and Ude, A. (2017). Enhancing the performance of adaptive iterative learning control with reinforcement learning. \textit{IEEE/RSJ Int. Conf. on IROS}, 2192-2199. 	

\bibitem[Oomen, 2018]{mikroniek} Oomen, T. (2018). Learning in machines: towards intelligent mechatronic systems through iterative control. \textit{Mikroniek}, 6, 5-11.

\bibitem[Oomen and Rojas, 2017]{oomen_rojas_2017} Oomen, T. and Rojas, C. (2017). Sparse iterative learning control with application to a wafer stage: achieving performance, resource efficiency, and task flexibility. \textit{Mechatronics}, 47, 134-147.

\bibitem[Phan and Frueh, 1996]{phan_frueh_1996} Phan, M. and Frueh, J. (1996). Learning control for trajectory tracking using basis functions. In \textit{Proc. of the 35th IEEE Conf. on Decision and Control}, 2490–2492.

\bibitem[Recht, 2019]{recht_2019} Recht, B. (2019). A tour of reinforcement learning: the view from continuous control. \textit{Annual Review of Control, Robotics, and Autonomous Systems}, 2, 253-279.


\bibitem[De Rozario and Oomen, 2019]{rozario_oomen_2019} de Rozario, R. and Oomen, T. (2019). Data-driven iterative inversion-based control: achieving robustness through nonlinear learning. \textit{Automatica}, 107, 342-352.

\bibitem[Sutton and Barto, 2017]{Book:RL} Sutton, R. and Barto, A. (2017). \textit{Reinforcement Learning: An Introduction.} MIT Press.

\bibitem[Sutton et al., 2000]{sutton_2000} Sutton, R., McAllester, D., Singh, S., and Mansour, Y. (2000). Policy gradient methods for reinforcement learning with function approximation. In \textit{Advances in Neural Information Processing Systems 12, MIT Press}, 1057-1063.


\bibitem[Turchetta et al., 2019]{turchetta} Turchetta, M., Krause, A., and Trimpe, S. (2019). Robust model-free reinforcement learning with multi-objective bayesian optimization. \textit{arXiv 1910.13399 cs.RO.}


\bibitem[Van de Wijdeven and Bosgra, 2010]{wijdeven_bosgra_2010} van de Wijdeven, J. and Bosgra, O. (2010). Using basis functions in iterative learning control: analysis and design theory. \textit{Int. Journal of Control}, 83(4), 661-675.

\bibitem[Van Zundert et al., 2016]{zundert_2016} van Zundert, J., Bolder, J., Koekebakker, S., and Oomen, T. (2016). Resource-efficient ILC for LTI/LTV systems through LQ tracking and stable inversion: enabling large feedforward tasks on a position-dependent printer. \textit{Mechatronics}, 38, 76-90.



%
%
%
%
%
%
%
%
%
%
%
%
%
%
%
%
%
%
%
%
%
%
%
%
%
%
%
%
%
%
%
%

\end{thebibliography}

\end{document}